\documentclass[aps,prd,two column,amsmath,amssymb,showpacs]{revtex4}
\usepackage{txfonts}
\usepackage{mathbbold}
\usepackage{epsfig,bm,dcolumn}
\usepackage{graphicx}
\usepackage{color}
\usepackage{overpic}

\newcommand{\feyn}[1]{\setbox0=\hbox{\ensuremath{#1}}\hbox to\wd0{\hbox to0pt{\hbox to\wd0{\hss/\hss}\hss}\box0}}

\begin{document}

\title{Ginzburg-Landau free energy of crystalline color superconductors: \\ A matrix formalism from solid-state physics }

\author{Gaoqing Cao$^{1}$ and Lianyi He$^{2}$\footnote{E-mail address: lianyi@lanl.gov; Present address: Department of Physics, Tsinghua University, Beijing 100084, China}}

\affiliation{1 Department of Physics, Tsinghua University and Collaborative Innovation Center of Quantum Matter, Beijing 100084, China\\
2 Theoretical Division, Los Alamos National Laboratory, Los Alamos, New Mexico 87545, USA}

\begin{abstract}
The Ginzburg-Landau (GL) free energy of crystalline color superconductors is important for understanding the nature of the phase transition to the normal quark matter and predicting the preferred crystal structure. So far the GL free energy at zero temperature has only been evaluated up to the sixth order in the condensate. To give quantitative reliable predictions we need to evaluate the higher-order terms. In this work, we present a new derivation of the GL free energy by using the discrete Bloch representation of the fermion field. This derivation introduces a simple matrix formalism without any momentum constraint, which may enable us to calculate the GL free energy to arbitrary order by using a computer.
\end{abstract}


\maketitle

\section{Introduction}\label{s1}

It is generally believed that the inhomogeneous Larkin-Ovchinnikov-Fulde-Ferrell (LOFF) phase appears in a superconductor when the pairing between different fermion species is under the circumstances of mismatched Fermi surfaces~\cite{LO1964,FF1964,Casalbuoni2004}. The mismatched Fermi surfaces are normally induced by the Zeeman energy splitting $2\delta\mu$ in a magnetic field~\cite{CC1962}. Early studies of the LOFF phase were restricted to 1D structures, which include the Fulde-Ferrell (FF) state~\cite{FF1964} with a plane-wave order parameter $\Delta(z)=\Delta e^{2iqz}$ and the Larkin-Ovhinnikov (LO) state~\cite{LO1964} with an antipodal-wave order parameter $\Delta(z)=2\Delta \cos(2qz)$. For $s$-wave pairing at weak coupling, it is known that the FF or LO state exists in a narrow window $\delta\mu_1<\delta\mu<\delta\mu_2$, where the lower critical field  $\delta\mu_1=0.707\Delta_0$ and the upper critical field $\delta\mu_2=0.754\Delta_0$~\cite{LO1964,FF1964} with $\Delta_0$ being the pairing gap at vanishing mismatch. However, since the thermodynamic critical field is much lower than $\delta\mu_1$ due to strong orbit effect, it is rather hard to observe the LOFF state in ordinary superconductors \cite{CC1962}. In recent years, experimental evidences for the LOFF state in some superconducting materials have been
reported~\cite{Heavyfermion,HighTc,Organic,FeSe}.

The studies of color superconductivity in dense quark matter promoted new interests in the LOFF state
\cite{Alford2001,Bowers2002,Shovkovy2003,Alford2004,EFG2004,Huang2004,Casalbuoni2005,Fukushima2005,Ren2005,Gorbar2006,Three-flavor,He2007, Anglani2014}.
Color superconductivity in dense quark matter appears due to the attractive interactions in certain diquark channels ~\cite{CSC01,CSC02,CSC03,CSC04,CSCreview}. Under the compact star constraints, different quark flavors ($u$, $d$, and $s$) acquire mismatched Fermi surfaces because of the Beta equilibrium and the electric charge neutrality. On the other hand, recent experiments on ultracold atomic Fermi gases provided a controllable way to study the fermion superfluidity with population imbalance~\cite{Atomexp,Sheehy2006}. Quark color superconductors under compact star constraints as well as atomic Fermi gases with population imbalance are rather clean systems to realize the long-sought LOFF phase.

In addition to the simple FF and LO states, there exist a variety of crystal structures. The general form of the crystal structure of the order parameter can be expressed as~\cite{Casalbuoni2004}
\begin{eqnarray}\label{crystal}
\Delta({\bf r})=\sum_{k=1}^P\Delta e^{2iq\hat{\bf n}_k\cdot{\bf r}}.
\end{eqnarray}
A specific crystal structure corresponds to a multi-wave configuration determined by the $P$ unit vectors ${\bf n}_{k}$ ($k=1,2,...,P$). Around the tricritical point in the temperature-mismatch phase diagram, the LOFF phase can be studied rigorously by using the Ginzburg-Laudau (GL) analysis since both the gap parameter $\Delta$ and the pair momentum $q$ are vanishingly small \cite{Casalbuoni2004}. It was found that the LO state is preferred  near the tricritical point~\cite{Buzdin1997,Combescot2002,Ye2007}. However, the real ground state of the LOFF phase is still not quite clear due to the limited theoretical approaches at zero temperature. Various theoretical approaches suggested that the LOFF state has a complicated crystal structure near its phase transition to the normal state~\cite{Bowers2002,EFG2004,Combescot2004,Combescot2005,Buballa2009,Cao2015}. A recent self-consistent treatment of the 1D modulation showed that a solitonic lattice structure is preferred near the phase transition to the BCS state~\cite{Buballa2009}.

The GL free energy is important for us to understand the nature of the phase transition to the normal state and to search for the most preferred crystal structure near the phase transition point. In a pioneer work, Bowers and Rajagopal investigated 23 crystal structures at weak coupling by using the GL approach~\cite{Bowers2002}. They evaluated the GL free energy up to the sixth order in $\Delta$,
\begin{eqnarray}
\frac{\Omega_{\rm GL}(\Delta)}{N_0}= P\alpha\Delta^2+\frac{1}{2}\beta\Delta^4+\frac{1}{3}\gamma\Delta^6+{\cal O}(\Delta^8),
\end{eqnarray}
where $N_0$ is the density of state at the Fermi surface. The coefficient $\alpha$ is universal for all crystal structures and is  given by~\cite{Bowers2002}
\begin{eqnarray}\label{alpha}
\alpha(\delta\mu,q)=-1+\frac{\delta\mu}{2q}\ln\left(\frac{q+\delta\mu}{q-\delta\mu}\right)
-\frac{1}{2}\ln\left[\frac{\Delta_0^2}{4(q^2-\delta\mu^2)}\right].
\end{eqnarray}
Near the conventional second-order phase transition point $\delta\mu=\delta\mu_2$ with the optimal pair momentum
$q=1.1997\delta\mu$, we have $\alpha\simeq(\delta\mu-\delta\mu_2)/\delta\mu_2$. The GL approach is meaningful if the phase transition is of second order or weak first order.

The GL coefficients $\beta$ and $\gamma$ for 23 crystal structures were evaluated in~\cite{Bowers2002}. For most of the structures, we find $\beta<0$, which leads to first-order phase transitions.  Among the structures with $\gamma>0$, the favored one seems to be the body-centered cubic (BCC) structure with $P=6$. The order parameter can be expressed as
\begin{equation}
\Delta({\bf r})=2\Delta\left[\cos(2qx)+\cos(2qy)+\cos(2qz)\right].
\end{equation}
Further, it was \emph{conjectured} that the face-centered cubic (FCC) structure with $P=8$ is the most preferred structure since both $\beta$ and $\gamma$ are negative and their absolute values are the largest~\cite{Bowers2002}. The order parameter can be expressed as
\begin{eqnarray}
\Delta({\bf r})=8\Delta\cos\left(\frac{2qx}{\sqrt{3}}\right)\cos\left(\frac{2qy}{\sqrt{3}}\right)\cos\left(\frac{2qz}{\sqrt{3}}\right).
\end{eqnarray}
For the BCC structure, the GL free energy up to the sixth order in $\Delta$ predicts a \emph{strong first-order} phase transition at $\delta\mu=\delta\mu_*\simeq3.6\Delta_0$ with a large gap parameter $\Delta\simeq0.8\Delta_0$ at $\delta\mu\lesssim\delta\mu_*$~\cite{Bowers2002}. The prediction of a strong first-order phase transition indicates that the GL approach is not valid or the higher-order terms in $\Delta$ are important. For the FCC structure, the GL free energy up to the sixth order in $\Delta$ actually gives \emph{no prediction} because both $\beta$ and $\gamma$ are negative.

On the other hand, there exist some other approaches that do not use the GL approximation. Combescot and Mora employed Eilenberger's quasiclassical equation with a Fourier expansion for the order parameter~\cite{Combescot2004,Combescot2005}. This approach predicted that the BCC-normal phase transition is of \emph{rather weak first order}: The upper critical field $\delta\mu_*$ is only $3.7\%$ higher than $\delta\mu_2$ and $\Delta\simeq0.1\Delta_0$ at $\delta\mu\lesssim\delta\mu_*$~\cite{Combescot2005}. For the FCC structure, it was found that its upper critical field is only $1.6\%$ higher than $\delta\mu_2$ and hence it is less favored than BCC.  Another study employed a solid-state-like approach which calculated the free energy of the BCC structure by directly diagonalizing the Hamiltonian matrix in the Bloch space~\cite{Cao2015}. This approach also predicted that the BCC-normal phase transition is of weak first order and the upper critical field of BCC is slightly higher than $\delta\mu_2$.

The contradiction between the predictions from the above approaches and from the GL approach indicates that the higher-order terms in the GL free energy are rather important for a quantitative prediction. Actually, for a first-order phase transition, the higher-order expansions are crucial to determine whether the first-order transition is weak or strong. For an intuitive understanding, let us add the eighth-order term in the GL free energy for the BCC structure. We have
\begin{eqnarray}
\frac{\Omega_{\rm GL}(\Delta)}{N_0}= P\alpha\Delta^2+\frac{1}{2}\beta\Delta^4+\frac{1}{3}\gamma\Delta^6+\frac{1}{4}\eta\Delta^8
+{\cal O}(\Delta^{10}).
\end{eqnarray}
As a naive example, we assume $\eta\geq0$. For $\eta=0$, we obtain a strong first-order phase transition. However, the first-order phase transition becomes weaker and weaker if we increase the value of $\eta$. For $\eta\rightarrow+\infty$, the phase transition approaches second order and $\delta\mu_*\rightarrow\delta\mu_2$. On the other hand, if $\eta$ is small or even negative, the higher-order terms such as
${\cal O}(\Delta^{10})$ may become important. Therefore, to give more precise predictions within the GL approach we need to evaluate the GL free energy to a sufficiently high order in $\Delta$. Actually, if we can evaluate the GL free energy to arbitrary order and determine the convergence properties of the GL series, we may even treat the strong first-order phase transition within the GL approach.

The higher-order GL coefficients can be evaluated by using the diagrammatic approach used in~\cite{Bowers2002}. In this approach, to evaluate the $2k$-th order GL coefficient one needs to sum all possible configurations that satisfy the momentum constraint
\begin{equation}\label{Mconstraint}
\sum_{i=1}^{2k}(-1)^{i+1}{\bf q}_i={\bf 0}
\end{equation}
for a set of $2k$ wave vectors $\{{\bf q}_1{\bf q}_2...{\bf q}_{2k}\}$ with ${\bf q}_i=q\hat{\bf n}_i$. For large $k$, the number of the configurations becomes very large for the crystal structures with large number of waves $P$. Moreover, one needs to introduce $2k$ Feynman parameters to evaluate the integrals. Therefore, the calculation of the higher-order terms is tedious and complicated in this formalism.

In this work, we introduce a new derivation of the GL free energy based on a solid-state physics approach to the crystal structures. This derivation provides a simple matrix formalism for the GL coefficients without any momentum constraint. We find that the formalism used in~\cite{Bowers2002} and our formalism can be attributed to two different \emph{representations} of the fermion field: The diagrammatic approach~\cite{Bowers2002} employs the
usual continuous momentum representation, while our derivation uses the discrete Bloch representation. Since the matrix operations can be easily realized by using a computer, this new formalism may enable us to calculate the GL free energy to arbitrary order in $\Delta$.

The paper is organized as follows. In Sec. \ref{s2} we briefly review the diagrammatic approach to the GL free energy. In Sec. \ref{s3} we present
our new derivation of the GL free energy by using the solid-state physics approach. The explicit matrix formalism of the GL coefficients for some crystal structures are presented in Sec. \ref{s4}. We summarize in Sec. \ref{s5}. The natural units $\hbar=k_{\rm B}=1$ will be used throughout.

\section{Continuous momentum representation}\label{s2}

We focus on the general two-flavor pairing at high density, low temperature, and weak coupling. In this case, the antiparticle degrees of freedom play no role. Therefore, we can start from a general effective Lagrangian for two-flavor pairing at high density. The Lagrangian density is
given by~\cite{Casalbuoni2004}
\begin{eqnarray}
{\cal L}_{\rm eff}=\psi^\dagger\left[i\partial_t-\varepsilon(\hat{\bf p})+\hat{\mu}\right]
\psi+\frac{g}{4}(\psi^\dagger\sigma_2\psi^*)(\psi^{\rm T}\sigma_2\psi),
\end{eqnarray}
where $\psi=(\psi_{\rm u},\psi_{\rm d})^{\rm T}$ denotes the two-flavor fermion field, $\varepsilon(\hat{\bf p})$ is the fermion dispersion with
$\hat{\bf p}=-i\mbox{\boldmath{$\nabla$}}$, $g$ is a contact coupling which represents the attractive interaction, and $\sigma_2$ is the second Pauli matrix in the flavor space. We shall neglect all other internal degrees of freedom, such as color and spin. These degrees of freedom contribute a simple degenerate factor and can be absorbed into the definition of the density of state $N_0$ at the Fermi surface. The fermion chemical potentials are specified by the diagonal matrix $\hat{\mu}={\rm diag}(\mu_{\rm u},\mu_{\rm d})$ in the flavor space, where
\begin{eqnarray}
\mu_{\rm u}=\mu+\delta\mu,\ \ \ \ \ \ \ \ \ \ \ \ \mu_{\rm d}=\mu-\delta\mu.
\end{eqnarray}
Here $\delta\mu$ plays the role of the mismatch between the Fermi surfaces. For ultra-relativistic fermion systems such as dense quark matter we take $\varepsilon(\bf p)=|{\bf p}|$. The model also works for nonrelativistic fermion systems such as cold atomic Fermi gases. In this case we take $\varepsilon({\bf p})={\bf p}^2$ (we set the fermion mass $M=1/2$ with proper units). Weak coupling requires that the pairing gap $\Delta_0$ at $\delta\mu=0$ is much smaller than the Fermi energy $\varepsilon_{\rm F}\simeq\mu$. At weak coupling, the physical results should be universal if we properly scale the physical quantities by using the pairing gap $\Delta_0$ and the density of states $N_0$ at the Fermi surface.

To evaluate the GL free energy, we start from the partition function ${\cal Z}$ of the system. In the imaginary time formalism, it is given by
\begin{eqnarray}
{\cal Z}=\int [d\psi][d\psi^\dagger]e^{-{\cal S}_{\rm eff}}
\end{eqnarray}
with the Euclidean action
\begin{eqnarray}
{\cal S}_{\rm eff}=-\int_0^{1/T}d\tau\int_Vd^3{\bf r}{\cal L}_{\rm eff}.
\end{eqnarray}
Here $\tau=it$ is the imaginary time and $T$ is the temperature. Once the partition function ${\cal Z}$ is evaluated, the free energy density is given by
\begin{eqnarray}
\Omega=-\frac{T}{V}\ln{\cal Z}.
\end{eqnarray}
Fermionic superconductivity is characterized by a nonzero expectation value of the difermion fields
\begin{eqnarray}
\varphi(\tau,{\bf r})=-g\psi_{\rm u}(\tau,{\bf r})\psi_{\rm d}(\tau,{\bf r})
\end{eqnarray}
The order parameter $\Delta({\bf r})=\langle\varphi(\tau,{\bf r})\rangle$ is static but inhomogeneous in the LOFF state.
At weak coupling and at low temperature, the order parameter fluctuation becomes negligible and we can employ the mean-field approach.
With the help of the Nambu-Gor'kov (NG) spinor
\begin{equation}
\Psi(\tau,{\bf r})=\left(\begin{array}{cc} \psi_{\rm u}^{\phantom{*}}(\tau,{\bf r}) \\ \psi_{\rm d}^*(\tau,{\bf r})\end{array}\right),
\end{equation}
the mean-field Lagrangian can be expressed as
\begin{eqnarray}
{\cal L}_{\rm{MF}}=\Psi^\dagger(\tau,{\bf r})(-\partial_\tau-{\cal H}_{\rm MF})\Psi(\tau,{\bf r})-\frac{|\Delta({\bf r})|^2}{g},
\end{eqnarray}
where the Hamiltonian operator is given by
\begin{eqnarray}
{\cal H}_{\rm MF}=\left(\begin{array}{cc} \varepsilon(\hat{\bf p})-\mu-\delta\mu & \Delta({\bf r})\\
\Delta^*({\bf r})& -\varepsilon(\hat{\bf p})+\mu-\delta\mu  \end{array}\right).
\end{eqnarray}

For convenience we first consider a finite system in a cubic box defined as $x,y,z\in [-L/2,L/2]$ and then set $L\rightarrow\infty$. Imposing the periodic boundary condition, the fermion momentum becomes discrete and is given by
\begin{eqnarray}\label{dmomentum}
{\bf p}=\frac{2\pi}{L}\left(l{\bf e}_x+m{\bf e}_y+n{\bf e}_z\right),\ \ \ l,m,n\in \mathbb{Z}.
\end{eqnarray}
To convert to the momentum space, we use the Fourier transformation for the fermion fields
\begin{eqnarray}\label{momentum-rep}
\Psi(\tau,{\bf r})&=&\frac{1}{\sqrt{V}}\sum_{{\bf p}}\sum_{\omega_n}\tilde{\Psi}(i\omega_n,{\bf p})
e^{-i\omega_n\tau+i{\bf p}\cdot{\bf r}},\nonumber\\
\Psi^\dagger(\tau,{\bf r})&=&\frac{1}{\sqrt{V}}\sum_{{\bf p}}\sum_{\omega_n}\tilde{\Psi}^\dagger(i\omega_n,{\bf p})
e^{i\omega_n\tau-i{\bf p}\cdot{\bf r}},
\end{eqnarray}
where $V=L^3$ is the system volume and $\omega_n=(2n+1)\pi T$ $(n\in \mathbb{Z})$ is the fermion Matsubara frequency.
For the general crystal structure of the order parameter given by (\ref{crystal}), the mean-field action in the momentum space can
be evaluated as
\begin{widetext}
\begin{eqnarray}
{\cal S}_{\rm{MF}}=\frac{V}{T}\frac{P\Delta^2}{g}-\frac{1}{T}\sum_{\omega_n,\omega_{n^\prime}}
\sum_{{\bf p},{\bf p}^\prime}\tilde{\Psi}^\dagger(i\omega_n,{\bf p})\left(i\omega_n\delta_{\omega_n,\omega_{n^\prime}}
\delta_{{\bf p},{\bf p}^\prime}-\delta_{\omega_n,\omega_{n^\prime}}{\cal H}_{{\bf p},{\bf p}^\prime}\right)
\tilde{\Psi}(i\omega_{n^\prime},{\bf p}^\prime),
\end{eqnarray}
\end{widetext}
where ${\bf p},{\bf p}^\prime$ are the discrete momenta given by (\ref{dmomentum}) and the Hamiltonian matrix ${\cal H}_{{\bf p},{\bf p}^\prime}$ reads
\begin{eqnarray}
{\cal H}_{{\bf p},{\bf p}^\prime}=\left(\begin{array}{cc} (\xi_{\bf p}-\delta\mu)\delta_{{\bf p},{\bf p}^\prime}
& \Delta (F_+)_{{\bf p},{\bf p}^\prime}\\ \Delta (F_-)_{{\bf p},{\bf p}^\prime}
& (-\xi_{\bf p}-\delta\mu)\delta_{{\bf p},{\bf p}^\prime}  \end{array}\right).
\end{eqnarray}
Here and in the following $\xi_{\bf p}=\varepsilon({\bf p})-\mu$. The matrices $F_\pm$ that characterizes the crystal structure are given by
\begin{eqnarray}
(F_\pm)_{{\bf p},{\bf p}^\prime}=\sum_{k=1}^P\delta_{{\bf p}-{\bf p}^\prime,\pm{\bf q}_k},
\end{eqnarray}
where ${\bf q}_k=q{\bf n}_k$.

The partition function in the mean-field approximation can be evaluated in the momentum representation. We have
\begin{eqnarray}
{\cal Z}_{\rm MF}=\int [d\tilde{\Psi}][d\tilde{\Psi}^\dagger]e^{-{\cal S}_{\rm MF}}.
\end{eqnarray}
Completing the Gaussian integral, we obtain the free energy
\begin{eqnarray}
\Omega=\frac{P}{g}\Delta^2-\frac{T}{V}\sum_{\omega_n}{\rm Trln}\left[\frac{(S^{-1})_{{\bf p},{\bf p}^\prime}}{T}\right],
\end{eqnarray}
where the Trln acts in the momentum space and the NG space. The inverse fermion propagator $S^{-1}$ defined in these spaces is given by
\begin{eqnarray}
(S^{-1})_{{\bf p},{\bf p}^\prime}(i\omega_n)=i\omega_n\delta_{{\bf p},{\bf p}^\prime}-{\cal H}_{{\bf p},{\bf p}^\prime}.
\end{eqnarray}
The free energy can be evaluated if we can diagonalize the Hamiltonian matrix ${\cal H}_{{\bf p},{\bf p}^\prime}$. However, this is infeasible because the momenta ${\bf p}$ and ${\bf p}^\prime$ become continuous in the large volume limit. We therefore turn to the GL expansion of the free energy. The standard field theoretical approach is to use the derivative expansion. First, we separate the``BCS" self-energy $\Sigma_\Delta$ and write
\begin{eqnarray}
(S^{-1})_{{\bf p},{\bf p}^\prime}=(S_0^{-1})_{{\bf p},{\bf p}^\prime}-(\Sigma_\Delta)_{{\bf p},{\bf p}^\prime},
\end{eqnarray}
where the inverse of the free fermion propagator reads
\begin{eqnarray}
(S_0^{-1})_{{\bf p},{\bf p}^\prime}=\left(\begin{array}{cc} (S_+^{-1})_{{\bf p},{\bf p}^\prime} & 0 \\ 0
& (S_{-}^{-1})_{{\bf p},{\bf p}^\prime}  \end{array}\right),
\end{eqnarray}
with the matrix elements given by
\begin{eqnarray}
(S_\pm^{-1})_{{\bf p},{\bf p}^\prime}=(i\omega_n+\delta\mu\mp\xi_{\bf p})\delta_{{\bf p},{\bf p}^\prime}.
\end{eqnarray}
The self-energy term $\Sigma_\Delta$ is given by
\begin{eqnarray}
(\Sigma_\Delta)_{{\bf p},{\bf p}^\prime}=\Delta F_{{\bf p},{\bf p}^\prime},
\end{eqnarray}
where the matrix $F$ reads
\begin{eqnarray}
F_{{\bf p},{\bf p}^\prime}=\left(\begin{array}{cc} 0 & (F_+)_{{\bf p},{\bf p}^\prime} \\ (F_-)_{{\bf p},{\bf p}^\prime}
& 0 \end{array}\right).
\end{eqnarray}
Using the derivative expansion, we obtain
\begin{eqnarray}
\Omega=\Omega_{\rm N}+\frac{P}{g}\Delta^2+\frac{T}{V}\sum_{\omega_n}
\sum_{l=1}^\infty\frac{\Delta^l}{l}{\rm Tr} \left[(S_0F)^l\right],
\end{eqnarray}
where $\Omega_{\rm N}$ is the free energy of the normal state,
\begin{eqnarray}\label{Normal}
\Omega_{\rm N}=-\frac{T}{V}\sum_{\omega_n}{\rm Trln}\left[\frac{(S_0^{-1})_{{\bf p},{\bf p}^\prime}}{T}\right].
\end{eqnarray}
To obtain the GL free energy, we first complete the trace in the NG space. It is easy to show the trace vanishes for odd $l$. After some manipulation, we obtain the GL free energy
\begin{eqnarray}
\Omega_{\rm GL}(\Delta)=\alpha_2\Delta^2+\sum_{k=2}^\infty\frac{\alpha_{2k}}{k}\Delta^{2k}.
\end{eqnarray}
The second-order GL coefficient $\alpha_2$ is given by
\begin{eqnarray}
\alpha_2=\frac{P}{g}+\frac{T}{V}\sum_{\omega_n}{\rm Tr} \left[S_+F_+S_-F_-\right].
\end{eqnarray}
The higher-order GL coefficients with $k\geq2$ are given by
\begin{eqnarray}
\alpha_{2k}=\frac{T}{V}\sum_{\omega_n}{\rm Tr} \left[(S_+F_+S_-F_-)^k\right].
\end{eqnarray}
Note that the trace is now taken only in the momentum space.

Next we complete the trace in the momentum space and obtain the integral expressions of the GL coefficients. For the second order, we have
\begin{widetext}
\begin{eqnarray}
{\rm Tr} \left[S_+F_+S_-F_-\right]&=&\sum_{a=1}^P\sum_{b=1}^P\sum_{\bf p}\sum_{{\bf p}_1,{\bf p}_2,{\bf p}_3}
\frac{\delta_{{\bf p},{\bf p}_1}}{i\omega_n+\delta\mu-\xi_{\bf p}}\delta_{{\bf p}_1-{\bf p}_2,2{\bf q}_a}
\frac{\delta_{{\bf p}_2,{\bf p}_3}}{i\omega_n+\delta\mu+\xi_{{\bf p}_2}}\delta_{{\bf p}_3-{\bf p},-2{\bf q}_b}\nonumber\\
&=&\sum_{a=1}^P\sum_{b=1}^P\sum_{\bf p}\sum_{{\bf p}^\prime}
\frac{1}{i\omega_n+\delta\mu-\xi_{\bf p}}\delta_{{\bf p}-{\bf p}^\prime,2{\bf q}_a}
\frac{1}{i\omega_n+\delta\mu+\xi_{{\bf p}^\prime}}\delta_{{\bf p}^\prime-{\bf p},-2{\bf q}_b}\nonumber\\
&=&\sum_{a=1}^P\sum_{\bf p}\frac{1}{i\omega_n+\delta\mu-\xi_{\bf p}}
\frac{1}{i\omega_n+\delta\mu+\xi_{{\bf p}-2{\bf q}_a}}.
\end{eqnarray}
Therefore, at zero temperature the second-order GL coefficient $\alpha_2$ can be expressed as
\begin{eqnarray}
\frac{\alpha_2}{P}=\frac{1}{g}+\int_{-\infty}^\infty\frac{dE}{2\pi}\int\frac{d^3{\bf p}}{(2\pi)^3}
\frac{1}{iE+\delta\mu-\xi_{\bf p}}\frac{1}{iE+\delta\mu+\xi_{{\bf p}-2{\bf q}}}.
\end{eqnarray}
We notice that $\alpha_2/P$ is universal for all crystal structures. The above integral suffers from ultraviolet (UV) divergence. In the weak coupling limit, the pairing and hence the momentum integral is dominated near the Fermi surface. It is convenient to use the regularization scheme that the momentum integral is restricted near the Fermi surface; i.e., $-\Lambda<|{\bf p}|-\mu<\Lambda$.
Using the fact that $\delta\mu, q\ll\Lambda\ll \mu$, we can express $\alpha_2$ as $\alpha_2=PN_0\alpha(\delta\mu,q)$, where
\begin{eqnarray}
\alpha(\delta\mu,q)=\frac{1}{gN_0}+\int_{-\infty}^\infty\frac{dE}{2\pi}\int_{-\Lambda}^\Lambda d\xi\int\frac{d\hat{\bf p}}{4\pi}
\frac{1}{iE+\delta\mu-\xi}\frac{1}{iE+\delta\mu+\xi-2\hat{\bf p}\cdot{\bf q}}
\end{eqnarray}
with $N_0=\mu^2/(2\pi^2)$ being the density of state at the Fermi surface and $\hat{\bf p}$ denoting the solid angle. The integrals can now be analytically worked out and the cutoff dependence can be  removed by using the pairing gap at vanishing mismatch,
$\Delta_0=2\Lambda e^{-1/(gN_0)}$. We finally obtain the analytical expression given by (\ref{alpha}).

For higher orders ($k\geq2$), the procedure is similar but becomes tedious. For example, for the fourth order ($k=2$), we have
\begin{eqnarray}
{\rm Tr} \left[(S_+F_+S_-F_-)^2\right]&=&\sum_{a=1}^P\sum_{b=1}^P\sum_{c=1}^P\sum_{d=1}^P\sum_{\bf p}\sum_{{\bf p}_1,{\bf p}_2,\ldots,{\bf p}_7}
\Bigg(\frac{\delta_{{\bf p},{\bf p}_1}}{i\omega_n+\delta\mu-\xi_{\bf p}}\delta_{{\bf p}_1-{\bf p}_2,2{\bf q}_a}
\frac{\delta_{{\bf p}_2,{\bf p}_3}}{i\omega_n+\delta\mu+\xi_{{\bf p}_2}}\delta_{{\bf p}_3-{\bf p}_4,-2{\bf q}_b}\nonumber\\
&&\ \ \ \ \ \ \ \ \ \ \ \ \ \ \ \ \ \ \ \ \ \ \ \ \ \ \ \ \ \ \ \ \ \ \ \ \ \ \ \ \ \ \ \
\frac{\delta_{{\bf p}_4,{\bf p}_5}}{i\omega_n+\delta\mu-\xi_{{\bf p}_4}}\delta_{{\bf p}_5-{\bf p}_6,2{\bf q}_c}
\frac{\delta_{{\bf p}_6,{\bf p}_7}}{i\omega_n+\delta\mu+\xi_{{\bf p}_6}}\delta_{{\bf p}_7-{\bf p},-2{\bf q}_d}\Bigg)
\nonumber\\
&=&\sum_{a=1}^P\sum_{b=1}^P\sum_{c=1}^P\sum_{d=1}^P\delta_{{\bf q}_a-{\bf q}_b+{\bf q}_c-{\bf q}_d,{\bf 0}}\sum_{\bf p}
\Bigg(\frac{1}{i\omega_n+\delta\mu-\xi_{\bf p}}\frac{1}{i\omega_n+\delta\mu+\xi_{{\bf p}-2{\bf q}_a}}\nonumber\\
&&\ \ \ \ \ \ \ \ \ \ \ \ \ \ \ \ \ \ \ \ \ \ \ \ \ \ \ \ \ \ \ \ \ \ \ \ \ \ \ \ \ \ \ \
\frac{1}{i\omega_n+\delta\mu-\xi_{{\bf p}-2{\bf q}_a+2{\bf q}_b}}
\frac{1}{i\omega_n+\delta\mu+\xi_{{\bf p}-2{\bf q}_a+2{\bf q}_b-2{\bf q}_c}}\Bigg).
\end{eqnarray}
At weak coupling, the $2k$-th order GL coefficient ($k\geq2$) can be generally expressed as
\begin{eqnarray}
\alpha_{2k}=N_0\sum_{{\bf q}_1,{\bf q}_2,\ldots,{\bf q}_{2k}}J_{2k}({\bf q}_1{\bf q}_2\ldots{\bf q}_{2k})\delta_{{\bf q}_s,{\bf 0}},\ \ \ \ \ \
{\bf q}_s=\sum_{i=1}^{2k}(-1)^{i+1}{\bf q}_i.
\end{eqnarray}
Here the summation over each ${\bf q}_i$ means the summation over all $P$ wave vectors ${\bf q}_a$ ($a=1,2,...,P$). The quantity
$J_{2k}({\bf q}_1{\bf q}_2\ldots{\bf q}_{2k})$ for general $k$ can be expressed in a compact form,
\begin{eqnarray}
J_{2k}({\bf q}_1{\bf q}_2\ldots{\bf q}_{2k})=\int_{-\infty}^\infty\frac{dE}{2\pi}\int_{-\infty}^\infty d\xi\int\frac{d\hat{\bf p}}{4\pi}
\prod_{i=1}^k\frac{1}{iE+\delta\mu-\xi+2\hat{\bf p}\cdot {\bf k}_i}\frac{1}{iE+\delta\mu+\xi-2\hat{\bf p}\cdot{\bf l}_i},
\end{eqnarray}
\end{widetext}
where the momenta ${\bf k}_i$ and ${\bf l}_i$ are given by (we define ${\bf q}_0={\bf 0}$ for convenience)
\begin{eqnarray}
{\bf k}_i=\sum_{n=0}^{2i-2}(-1)^{n+1}{\bf q}_n,\ \ \ \ \ \ \ \ {\bf l}_i=\sum_{n=0}^{2i-1}(-1)^{n+1}{\bf q}_n.
\end{eqnarray}
Note that we have set $\Lambda\rightarrow\infty$ since the integral over $\xi$ is free from UV divergence for $k\geq2$.

The above formalism is completely the same as that obtained by using the diagrammatic approach~\cite{Bowers2002}. For the nonrelativistic case, the results can be obtained
by replacing the pair momentum $q$ with $v_{\rm F}q$ where $v_{\rm F}=2\sqrt{\mu}$ is the Fermi velocity. The density of state $N_0$ at the Fermi surface is replaced by $N_0=\sqrt{\mu}/(4\pi^2)$. Therefore, at weak coupling, the GL free energy is universal for both the relativistic case and the nonrelativistic case once we properly express the free energy in terms of the density of state $N_0$ at the Fermi surface, the pairing gap $\Delta_0$ at vanishing mismatch, and the quantity $v_{\rm F}q$ ($v_{\rm F}=1$ for the ultra-relativistic case).

The GL coefficients up to the sixth order ($k\leq3$) for 23 crystal structures have been evaluated numerically by Bowers and Rajagopal~\cite{Bowers2002}. They introduced Feynman parameters to evaluate the integral $J_{2k}({\bf q}_1{\bf q}_2\ldots{\bf q}_{2k})$. The advantage of the above formalism obtained by using the continuous momentum representation is that one can directly approach the weak coupling limit by using the momentum cutoff scheme $-\Lambda<|{\bf p}|-\mu<\Lambda$. However, for higher-order coefficients with large $k$, the calculation becomes complicated and tedious. First, for general $k$, one needs to introduce $2k$ Feynman parameters $x_1,x_2,...,x_k$ and $y_1,y_2,...,y_k$. At large $k$, the integral over the Feynman parameters becomes complicated. Second, to obtain the GL coefficients, one needs to sum over all possible configurations that satisfy the constraint $\sum_{i=1}^{2k}(-1)^{i+1}{\bf q}_i={\bf 0}$. At large $k$, the number of these configurations becomes also large, which makes the calculation tedious. To the best of our knowledge, no results of the higher-order GL coefficients ($k\geq4$) have been reported so far.

\section{Solid-State Physics Approach: Discrete representation}\label{s3}

In this section, we turn to a discrete representation inspired by solid-state physics. For a specific crystal structure given by (\ref{crystal}), it is periodic in coordinate space. The order parameter can be alternatively expressed as
\begin{eqnarray}
\Delta({\bf r})=\Delta f({\bf r}),
\end{eqnarray}
where
\begin{equation}
f({\bf r})=\sum_{k=1}^Pe^{2iq\hat{\bf n}_k\cdot{\bf r}}
\end{equation}
is a periodic function. Here we assume that the function $f({\bf r})$ corresponds to a 3D lattice structure. The derivation of the GL free energy can be easily generalized to 1D and 2D lattice structures. For a 3D lattice structure, the unit cell is generated by three linearly independent vectors ${\bf a}_1$, ${\bf a}_2$, and ${\bf a}_3$. We can accordingly define the reciprocal space generated by three linearly independent vectors
${\bf b}_1$, ${\bf b}_2$, and ${\bf b}_3$ with ${\bf b}_i$ obtained by the relation ${\bf a}_i\cdot{\bf b}_j=2\pi\delta_{ij}$. The periodicity of the order parameter means $f({\bf r})=f({\bf r}+{\bf a}_i)$. Therefore, the function $f({\bf r})$ can be decomposed into a discrete set of Fourier components. We have
\begin{eqnarray}\label{Fourier}
f({\bf r})&=&\sum_{{\bf G}}f_{\bf G}e^{i{\bf G}\cdot {\bf r}}=\sum_{l,m,n=-\infty}^\infty f_{lmn}e^{i{\bf G}_{lmn}\cdot {\bf r}},\nonumber\\
f^*({\bf r})&=&\sum_{{\bf G}}f_{\bf G}^*e^{-i{\bf G}\cdot {\bf r}}=\sum_{l,m,n=-\infty}^\infty f_{lmn}^*e^{-i{\bf G}_{lmn}\cdot {\bf r}},
\end{eqnarray}
where the reciprocal lattice vector ${\bf G}$ is given by
\begin{eqnarray}
{\bf G}={\bf G}_{lmn}=l{\bf b}_1+m{\bf b}_2+n{\bf b}_3,\ \ \ l,m,n\in \mathbb{Z}.
\end{eqnarray}
The Fourier components $f_{\bf G}$ and $f^*_{\bf G}$ are given by
\begin{eqnarray}
f_{\bf G}&=&\frac{1}{V_c}\int_c d^3{\bf r}f({\bf r})e^{-i{\bf G}\cdot{\bf r}},\nonumber\\
f_{\bf G}^*&=&\frac{1}{V_c}\int_c d^3{\bf r}f^*({\bf r})e^{i{\bf G}\cdot{\bf r}},
\end{eqnarray}
where $V_c$ is the volume of the unit cell and the integration is restricted in the cell volume. It is easy to show that
\begin{equation}
\sum_{\bf G}|f_{\bf G}|^2=P.
\end{equation}

Since the order parameter or pair potential $\Delta({\bf r})$ is periodic, the eigenvalue equation for the fermionic excitation spectrum, which is known as the Bogoliubov-de Gennes (BdG) equation, is analogous to the Schr\"{o}dinger equation of a quantum particle moving in a periodic potential. The BdG equation for the present system can be expressed as
\begin{eqnarray}
\left(\begin{array}{cc} \varepsilon(\hat{\bf p})-\mu-\delta\mu & \Delta({\bf r})\\
\Delta^*({\bf r})& -\varepsilon(\hat{\bf p})+\mu-\delta\mu  \end{array}\right)\phi_\lambda({\bf r})
=E_\lambda\phi_\lambda({\bf r}).
\end{eqnarray}
According to the Bloch theorem, the eigenfunction $\phi_\lambda({\bf r})$ takes the form of the Bloch function; i.e.,
\begin{eqnarray}
\phi_\lambda({\bf r})=e^{i{\bf k}\cdot{\bf r}}\phi_{\lambda{\bf k}}({\bf r}),
\end{eqnarray}
where the lattice momentum ${\bf k}$ is restricted in the Brillouin zone (BZ) and the function $\phi_{\lambda{\bf k}}({\bf r})$ has the same periodicity as the order parameter $\Delta({\bf r})$. We therefore have
the similar Fourier expansion
\begin{eqnarray}
\phi_{\lambda{\bf k}}({\bf r})=\sum_{\bf G}\phi_{\bf G}e^{i{\bf G}\cdot {\bf r}}
=\sum_{l,m,n=-\infty}^\infty \phi_{lmn}e^{i{\bf G}_{lmn}\cdot {\bf r}}.
\end{eqnarray}
Substituting this expansion into the BdG equation, we finally obtain a matrix equation in the ${\bf G}$-space,
\begin{eqnarray}\label{matrixEq}
\sum_{{\bf G}^\prime}{\cal H}_{{\bf G},{\bf G}^\prime}({\bf k})\phi_{{\bf G}^\prime}=E_\lambda({\bf k}) \phi_{\bf G},
\end{eqnarray}
where the Hamiltonian matrix ${\cal H}_{{\bf G},{\bf G}^\prime}({\bf k})$ is given by
\begin{eqnarray}\label{Hmatrix}
{\cal H}_{{\bf G},{\bf G}^\prime}({\bf k})=\left(\begin{array}{cc} (\xi_{{\bf k}+{\bf G}}-\delta\mu)\delta_{{\bf G},{\bf G}^\prime}
& \Delta f_{{\bf G}-{\bf G}^\prime} \\ \Delta f^*_{{\bf G}^\prime-{\bf G}}
& (-\xi_{{\bf k}+{\bf G}}-\delta\mu)\delta_{{\bf G},{\bf G}^\prime}  \end{array}\right).
\end{eqnarray}
For a given momentum ${\bf k}$ in the BZ, we can solve the eigenvalues $E_{\lambda}({\bf k})$ by diagonalizing the above
matrix in the discrete ${\bf G}$-space.  This means that the fermionic excitation spectrum forms a band structure, in analogy to
the energy spectrum of a quantum particle moving in a periodic potential.

Now we turn to the field theory. We consider a finite system spanned by three vectors $N_1{\bf a}_1$, $N_2{\bf a}_2$, and $N_3{\bf a}_3$ and assume periodic boundary condition. Then the system contains $N_1N_2N_3$ unit cells and the thermodynamic limit can be reached by setting $N_i\rightarrow\infty$. In accordance with the Fourier expansion (\ref{Fourier}) and the matrix equation (\ref{matrixEq}), we expand the
fermion field $\Psi(\tau,{\bf r})$ in terms of the Bloch function rather than using the usual momentum representation (\ref{momentum-rep}).
We write
\begin{widetext}
\begin{eqnarray}
\Psi(\tau,{\bf r})&=&\frac{1}{\sqrt{V}}\sum_{{\bf k}\in{\rm BZ}}e^{i{\bf k}\cdot{\bf r}}\sum_{\omega_n}\sum_{{\bf G}}
\tilde{\Psi}(i\omega_n,{\bf k},{\bf G})e^{-i\omega_n\tau+i{\bf G}\cdot{\bf r}},\nonumber\\
\Psi^\dagger(\tau,{\bf r})&=&\frac{1}{\sqrt{V}}\sum_{{\bf k}\in{\rm BZ}}e^{-i{\bf k}\cdot{\bf r}}\sum_{\omega_n}\sum_{{\bf G}}
\tilde{\Psi}^\dagger(i\omega_n,{\bf k},{\bf G})e^{i\omega_n\tau-i{\bf G}\cdot{\bf r}}.
\end{eqnarray}
One can recover the usual momentum representation (\ref{momentum-rep}) by using the fact that ${\bf k}+{\bf G}$ can generate all possible momentum ${\bf p}$ in the momentum space. This expansion defines a new representation with two different quantum numbers ${\bf k}$ and ${\bf G}$. We call this Bloch representation. In the Bloch representation, the mean-field action can be evaluated as
\begin{eqnarray}
{\cal S}_{\rm{MF}}=\frac{V}{T}\frac{\Delta^2}{g}\sum_{\bf G}|f_{\bf G}|^2-\frac{1}{T}\sum_{\omega_n,\omega_{n^\prime}}
\sum_{{\bf k},{\bf k}^\prime\in{\rm BZ}}\sum_{{\bf G},{\bf G}^\prime}\tilde{\Psi}^\dagger(i\omega_n,{\bf k},{\bf G})
\left[i\omega_n\delta_{\omega_n,\omega_{n^\prime}}\delta_{{\bf k},{\bf k}^\prime}\delta_{{\bf G},{\bf G}^\prime}
-\delta_{\omega_n,\omega_{n^\prime}}\delta_{{\bf k},{\bf k}^\prime}{\cal H}_{{\bf G},{\bf G}^\prime}({\bf k})\right]
\tilde{\Psi}(i\omega_{n^\prime},{\bf k}^\prime,{\bf G}^\prime),
\end{eqnarray}
\end{widetext}
where the Hamiltonian matrix ${\cal H}_{{\bf G},{\bf G}^\prime}({\bf k})$ is given by (\ref{Hmatrix}). In mathematics, the Bloch representation corresponds to a similarity transformation of the usual momentum representation, which makes the Hamiltonian matrix
${\cal H}_{{\bf p},{\bf p}^\prime}$ block diagonal with the blocks characterized by the lattice momentum ${\bf k}$. The functional path integral can be worked out by performing integrals over $\tilde{\Psi}^\dagger(i\omega_n,{\bf k},{\bf G})$ and $\tilde{\Psi}(i\omega_{n^\prime},{\bf k}^\prime,{\bf G}^\prime)$ for all possible values of $\{i\omega_n,{\bf k},{\bf G}\}$ and $\{i\omega_{n^\prime},{\bf k}^\prime,{\bf G}^\prime\}$. Since the inverse fermion propagator is  diagonal in the frequency space and the ${\bf k}$-space, the free energy can be expressed as
\begin{eqnarray}\label{free-energy-G}
\Omega=\frac{P}{g}\Delta^2-\frac{T}{V}\sum_{\omega_n}\sum_{{\bf k}\in{\rm BZ}}
{\rm Trln}\left[\frac{(S^{-1})_{{\bf G},{\bf G}^\prime}}{T}\right],
\end{eqnarray}
where the Trln acts in the ${\bf G}$-space and the NG space. The inverse fermion propagator $S^{-1}$ defined
in the ${\bf G}$-space and the NG space is given by
\begin{eqnarray}
(S^{-1})_{{\bf G},{\bf G}^\prime}(i\omega_n,{\bf k})=i\omega_n\delta_{{\bf G},{\bf G}^\prime}-{\cal H}_{{\bf G},{\bf G}^\prime}({\bf k}).
\end{eqnarray}
The free energy can be evaluated if we can diagonalize the Hamiltonian matrix ${\cal H}_{{\bf G},{\bf G}^\prime}({\bf k})$ to obtain
all the eigenvalues or the band spectrum $\{E_{\lambda}({\bf k})\}$. This is in principle feasible because the ${\bf G}$-space is discrete.
Since the matrix has infinite dimensions, we should make a truncation $-D\leq l,m,n\leq D$ ($D\in \mathbb{Z}^+$) to perform the diagonalization. The size of the matrix we need to diagonalize is $2(2D+1)^3$. To achieve convergence to the limit $D\rightarrow\infty$ we normally need a large cutoff $D$, which leads to a large computing cost~\cite{Cao2015}.

Then we turn to the GL expansion based on the expression (\ref{free-energy-G}) of the free energy. The procedure is the same as we used in Sec. II. First, for the inverse fermion propagator $S^{-1}$, we separate the BCS self-energy and obtain
\begin{eqnarray}
(S^{-1})_{{\bf G},{\bf G}^\prime}=(S_0^{-1})_{{\bf G},{\bf G}^\prime}-(\Sigma_\Delta)_{{\bf G},{\bf G}^\prime}.
\end{eqnarray}
Here the inverse of the free fermion propagator reads
\begin{eqnarray}
(S_0^{-1})_{{\bf G},{\bf G}^\prime}=\left(\begin{array}{cc} (S_+^{-1})_{{\bf G},{\bf G}^\prime} & 0 \\ 0
& (S_{-}^{-1})_{{\bf G},{\bf G}^\prime}  \end{array}\right)
\end{eqnarray}
with the matrix elements given by
\begin{eqnarray}
(S_\pm^{-1})_{{\bf G},{\bf G}^\prime}=(i\omega_n+\delta\mu\mp\xi_{{\bf k}+{\bf G}})\delta_{{\bf G},{\bf G}^\prime}.
\end{eqnarray}
The self-energy term $\Sigma_\Delta$ can be expressed as
\begin{equation}
(\Sigma_\Delta)_{{\bf G},{\bf G}^\prime}=\Delta F_{{\bf G},{\bf G}^\prime},
\end{equation}
where the matrix $F$ is defined as
\begin{eqnarray}
F_{{\bf G},{\bf G}^\prime}=\left(\begin{array}{cc} 0 & (F_+)_{{\bf G},{\bf G}^\prime} \\ (F_-)_{{\bf G},{\bf G}^\prime}
& 0 \end{array}\right).
\end{eqnarray}
Here the blocks $F_\pm$ defined in the ${\bf G}$-space are given by
\begin{eqnarray}
(F_+)_{{\bf G},{\bf G}^\prime}=f_{{\bf G}-{\bf G}^\prime},\ \ \ \ \ \ (F_-)_{{\bf G},{\bf G}^\prime}=f^*_{{\bf G}^\prime-{\bf G}}.
\end{eqnarray}
Using the derivative expansion, we obtain
\begin{eqnarray}
\Omega=\Omega_{\rm N}+\frac{P}{g}\Delta^2+\frac{T}{V}\sum_{\omega_n}\sum_{{\bf k}\in{\rm BZ}}
\sum_{l=1}^\infty\frac{\Delta^l}{l}{\rm Tr}\left[(S_0F)^l\right],
\end{eqnarray}
where $\Omega_{\rm N}$ is the free energy of the normal state,
\begin{eqnarray}
\Omega_{\rm N}=-\frac{T}{V}\sum_{\omega_n}\sum_{{\bf k}\in{\rm BZ}}{\rm Trln}\left[\frac{(S_0^{-1})_{{\bf G},{\bf G}^\prime}}{T}\right].
\end{eqnarray}
Using the fact that $S_0^{-1}$ is diagonal and ${\bf k}+{\bf G}$ generates all continuum momenta ${\bf p}$, we recover the usual expression (\ref{Normal}).

Completing the trace in the NG space, we find that the trace vanishes for odd $l$. Then the GL free energy can be expressed as
\begin{eqnarray}
\Omega_{\rm GL}=\alpha_2\Delta^2+\sum_{k=2}^\infty\frac{\alpha_{2k}}{k}\Delta^{2k}.
\end{eqnarray}
The second-order GL coefficient is given by
\begin{eqnarray}
\alpha_2=\frac{P}{g}+\frac{T}{V}\sum_{\omega_n}\sum_{{\bf k}\in{\rm BZ}}{\rm Tr}(S_+F_+S_-F_-).
\end{eqnarray}
The higher-order GL coefficients $\alpha_{2k}$ ($k\geq2$) read
\begin{eqnarray}
\alpha_{2k}=\frac{T}{V}\sum_{\omega_n}\sum_{{\bf k}\in{\rm BZ}}{\rm Tr}\left[(S_+F_+S_-F_-)^k\right].
\end{eqnarray}
Note that the trace is now taken only in the ${\bf G}$-space. At zero temperature and in the large volume limit, we obtain
\begin{eqnarray}
\alpha_2=\frac{P}{g}+\int_{-\infty}^\infty\frac{dE}{2\pi}\int_{\rm BZ}\frac{d^3{\bf k}}{(2\pi)^3}{\cal A}_2(iE,{\bf k}).
\end{eqnarray}
and
\begin{eqnarray}
\alpha_{2k}=\int_{-\infty}^\infty\frac{dE}{2\pi}\int_{\rm BZ}\frac{d^3{\bf k}}{(2\pi)^3}{\cal A}_{2k}(iE,{\bf k}).
\end{eqnarray}
for $k\geq2$. Here the quantity ${\cal A}_{2k}(iE,{\bf k})$ is defined as
\begin{eqnarray}
{\cal A}_{2k}(iE,{\bf k})={\rm Tr}\ \Big\{\left[S_+(iE,{\bf k})F_+S_-(iE,{\bf k})F_-\right]^k\Big\}
\end{eqnarray}
with
\begin{eqnarray}
(S_\pm)_{{\bf G},{\bf G}^\prime}(iE,{\bf k})=\frac{\delta_{{\bf G},{\bf G}^\prime}}{iE+\delta\mu\mp\xi_{{\bf k}+{\bf G}}}.
\end{eqnarray}

The GL coefficients should be independent of the representation. Therefore, $\alpha_2/P$ is still universal for all crystal structure. We can evaluate it from the FF state. We have
\begin{equation}
\alpha_2=P\alpha_2^{\rm FF}.
\end{equation}
For the FF state we have $f({\bf r})=f(z)=e^{2iqz}$, which can be regarded as a 1D crystal structure with periodicity $a=\pi/q$. We have the Fourier transformations
\begin{eqnarray}
f(z)=\sum_{n=-\infty}^\infty f_ne^{i\frac{2\pi n}{a}z},\ \ \ \ \ f^*(z)=\sum_{n=-\infty}^\infty f_n^*e^{-i\frac{2\pi n}{a}z},
\end{eqnarray}
where the Fourier components $f_n$ and $f_n^*$ are given by
\begin{eqnarray}
f_n=f_n^*=\delta_{n,1}.
\end{eqnarray}
Therefore, the matrices $F_\pm$ read
\begin{eqnarray}
(F_+)_{n,n^\prime}=\delta_{n-n^\prime,1},\ \ \ \ (F_-)_{n,n^\prime}=\delta_{n^\prime-n,1}.
\end{eqnarray}
The GL coefficient $\alpha_2^{\rm FF}$ can be expressed as
\begin{eqnarray}
\alpha_2^{\rm FF}=\frac{1}{g}+\int_{-\infty}^\infty\frac{dE}{2\pi}\int_0^\infty\frac{k_\perp dk_\perp}{2\pi}
\int_{-q}^q\frac{dk_z}{2\pi}{\cal A}_{2}(iE,k_\perp,k_z).
\end{eqnarray}
The trace in ${\cal A}_2$ can be worked out analytically. We have
\begin{eqnarray}
&&{\cal A}_{2}(iE,k_\perp,k_z)={\rm Tr} \left[S_+F_+S_-F_-\right]\nonumber\\
&=&\sum_{n}\sum_{n_1,n_2,n_3}
\frac{\delta_{n,n_1}}{iE+\delta\mu-\xi_{n}}\delta_{n_1-n_2,1}
\frac{\delta_{n_2,n_3}}{iE+\delta\mu+\xi_{n_2}}\delta_{n-n_3,1}\nonumber\\
&=&\sum_{n}\frac{1}{iE+\delta\mu-\xi_n}
\frac{1}{iE+\delta\mu+\xi_{n-1}},
\end{eqnarray}
where $\xi_n$ is defined as
\begin{eqnarray}
\xi_n(k_\perp,k_z)=\left[k_\perp^2+(k_z+2nq)^2\right]^\nu-\mu.
\end{eqnarray}
Here $\nu=1/2$ for the ultra-relativistic case and $\nu=1$ for the nonrelativistic case. Since $k_z+2nq$ generates all continuous momenta in the
$z$ direction, the final result for $\alpha_2$ can be expressed as (36).

For higher-order GL coefficients $\alpha_{2k}$ ($k\geq2$), the trace in the quantity ${\cal A}_{2k}(iE,{\bf k})$ becomes tedious.
However, because the ${\bf G}$-space is discrete, this procedure can be realized by using a computer with a proper code. Since the ${\bf G}$-space has infinite dimensions, we first make a truncation $-D\leq l,m,n\leq D$ to perform the matrix operations. Precise results can be approached by using a sufficiently large $D$. Some of the GL coefficients may suffer from UV divergence and a proper regularization scheme should be implemented. For ultra-relativistic dispersion $\varepsilon({\bf p})=|{\bf p}|$, $\alpha_2$ and $\alpha_4$ are divergent. For non-relativistic case with $\varepsilon({\bf p})=|{\bf p}|^2$, only $\alpha_2$ is divergent. The GL free energy can be expressed as
\begin{eqnarray}
\frac{\Omega_{\rm GL}(\Delta)}{N_0\delta\mu^2}=P\bar{\alpha}\left(\frac{\Delta}{\delta\mu}\right)^2+
\sum_{k=2}^\infty\frac{\bar{\alpha}_{2k}}{k}\left(\frac{\Delta}{\delta\mu}\right)^{2k},
\end{eqnarray}
where the GL coefficients become dimensionless. We have
\begin{equation}
\bar{\alpha}_{2k}=\frac{\delta\mu^{2k-2}}{N_0}\alpha_{2k}.
\end{equation}
These dimensionless GL coefficients are \emph{universal} in the weak coupling limit. Therefore, we can evaluate them by using the non-relativistic dispersion because all the coefficients $\alpha_{2k}$ ($k\geq2$) are free from UV divergence. On the other hand, in this matrix formalism, we need to treat the momenta ${\bf k}$ and ${\bf G}$ separately. Therefore, the usual momentum cutoff scheme $-\Lambda<\xi_{\bf p}<\Lambda$ is not proper to achieve the weak coupling limit. The dependence on the chemical potential $\mu$ becomes explicit in the matrix formalism. Weak coupling corresponds to the case $\mu\simeq\varepsilon_{\rm F}\gg\delta\mu$. In practice, we can vary the value of $\mu/\delta\mu$ and approach the weak coupling limit.

\section{Formalism for some crystal structures}\label{s4}

In the final part of this paper, we consider some crystal structures of which the order parameters $\Delta({\bf r})$ are real. We have $f_{\bf G}^*=f_{-{\bf G}}^{\phantom{*}}$ and hence
\begin{eqnarray}
(F_+)_{{\bf G},{\bf G}^\prime}=(F_-)_{{\bf G},{\bf G}^\prime}\equiv F_{{\bf G},{\bf G}^\prime}=f_{{\bf G}-{\bf G}^\prime}.
\end{eqnarray}
In the following, we list the explicit forms of the matrix $F$, the propagators $S_\pm$, and the GL coefficients $\alpha_{2k}$ ($k\geq2$). As we pointed out in Sec. \ref{s3}, we shall use the nonrelativistic dispersion $\varepsilon({\bf p})={\bf p}^2$. In practice, this leads to faster convergence and hence is better for numerical calculations.

\subsection{LO}

The LO state is a superposition of two antipodal plane waves ($P=2$) with
\begin{equation}
\hat{\bf n}_1=(0,0,1),\ \ \ \ \hat{\bf n}_2=(0,0,-1).
\end{equation}
The function $f({\bf r})$ can be expressed as
\begin{eqnarray}
f({\bf r})=f(z)=2\cos(2qz),
\end{eqnarray}
which forms a 1D crystal structure with periodicity $a=\pi/q$. The Fourier decomposition is given by
\begin{eqnarray}
f(z)=\sum_{n=-\infty}^\infty f_ne^{2niqz},
\end{eqnarray}
where the Fourier component $f_n$ reads
\begin{eqnarray}
f_n=\delta_{n,1}+\delta_{n,-1}.
\end{eqnarray}
The matrix form of the BdG equation is given by
\begin{eqnarray}
\sum_{n^\prime}{\cal H}_{n,n^\prime}({\bf k})\phi_{n^\prime}({\bf k})=E_\lambda({\bf k}) \phi_{n}({\bf k}),
\end{eqnarray}
where the Hamiltonian matrix ${\cal H}_{n,n^\prime}({\bf k})$ reads
\begin{eqnarray}
\left(\begin{array}{cc} (\xi_n-\delta\mu)\delta_{n,n^\prime} &
\Delta f_{n-n^\prime} \\ \Delta f_{n-n^\prime}
& (-\xi_n-\delta\mu)\delta_{n,n^\prime}  \end{array}\right)
\end{eqnarray}
with
\begin{eqnarray}
\xi_n(k_\perp,k_z)=k_\perp^2+(k_z+2nq)^2-\mu.
\end{eqnarray}
The BZ can be defined as $k_z\in[-q,q]$. The GL coefficients $\alpha_{2k}$ ($k\geq2$) can be expressed as
\begin{eqnarray}
\alpha_{2k}=\int_{-\infty}^\infty\frac{dE}{2\pi}\int_0^\infty\frac{k_\perp dk_\perp}{2\pi}
\int_{-q}^q\frac{dk_z}{2\pi}{\rm Tr}_{[n]}\left[(S_+FS_-F)^k\right].
\end{eqnarray}
The matrix elements of $S_\pm$ and $F$ are given by
\begin{eqnarray}
&&(S_\pm)_{n,n^\prime}=\frac{\delta_{n,n^\prime}}{iE+\delta\mu\mp\xi_n},\nonumber\\
&&F_{n,n^\prime}=\delta_{n,n^\prime+1}+\delta_{n,n^\prime-1}.
\end{eqnarray}

\subsection{Square}

The Square state is a superposition of four plane waves ($P=4$) with
\begin{eqnarray}
&&\hat{\bf n}_1=(1,0,0),\ \ \ \ \hat{\bf n}_2=(-1,0,0),\nonumber\\
&&\hat{\bf n}_3=(0,1,0),\ \ \ \ \hat{\bf n}_4=(0,-1,0).
\end{eqnarray}
The function $f({\bf r})$ can be expressed as
\begin{eqnarray}
f({\bf r})=f(x,y)=2\left[\cos(2qx)+\cos(2qy)\right],
\end{eqnarray}
which forms a 2D crystal structure. The unit cell is generated by two linearly independent vectors ${\bf a}_1=a{\bf e}_x$ and ${\bf a}_2=a{\bf e}_y$ with the lattice spacing $a=\pi/q$. The Fourier decomposition is given by
\begin{eqnarray}
f(x,y)=\sum_{l,m=-\infty}^\infty f_{lm}e^{i\frac{2\pi}{a}(lx+my)},
\end{eqnarray}
where the Fourier component $f_{lm}$ reads
\begin{eqnarray}
f_{lm}=(\delta_{l,1}+\delta_{l,-1})\delta_{m,0}+\delta_{l,0}(\delta_{m,1}+\delta_{m,-1}).
\end{eqnarray}
The matrix form of the BdG equation is given by
\begin{eqnarray}
\sum_{l^\prime m^\prime}{\cal H}_{[lm],[l^\prime m^\prime]}({\bf k})\phi_{[l^\prime m^\prime]}({\bf k})=E_\lambda({\bf k}) \phi_{[lm]}({\bf k}),
\end{eqnarray}
where the Hamiltonian matrix ${\cal H}_{[lm],[l^\prime m^\prime]}({\bf k})$ reads
\begin{eqnarray}
\left(\begin{array}{cc} (\xi_{lm}-\delta\mu)\delta_{l,l^\prime}\delta_{m,m^\prime} &
\Delta f_{l-l^\prime,m-m^\prime} \\ \Delta f_{l-l^\prime,m-m^\prime}
& (-\xi_{lm}-\delta\mu)\delta_{l,l^\prime}\delta_{m,m^\prime}  \end{array}\right)
\end{eqnarray}
with
\begin{eqnarray}
\xi_{lm}(k_x,k_y,k_z)=(k_x+2lq)^2+(k_y+2mq)^2+k_z^2-\mu.
\end{eqnarray}
The BZ can be defined as $k_x,k_y\in[-q,q]$. The GL coefficients $\alpha_{2k}$ ($k\geq2$) are given by
\begin{eqnarray}
\alpha_{2k}&=&\int_{-\infty}^\infty\frac{dE}{2\pi}\int_{-q}^q\frac{dk_x}{2\pi}\int_{-q}^q\frac{dk_y}{2\pi}
\int_{-\infty}^\infty\frac{dk_z}{2\pi}\nonumber\\
&&{\rm Tr}_{[lm]} \left[(S_+FS_-F)^k\right].
\end{eqnarray}
The matrix elements of $S_\pm$ and $F$ read
\begin{eqnarray}
&&(S_\pm)_{[lm],[l^\prime m^\prime]}=\frac{\delta_{l,l^\prime}\delta_{m,m^\prime}}{iE+\delta\mu\mp\xi_{lm}},\nonumber\\
&&F_{[lm],[l^\prime m^\prime]}=(\delta_{l,l^\prime+1}+\delta_{l,l^\prime-1})\delta_{m,m^\prime}\nonumber\\
&&\ \ \ \ \ \ \ \ \ \ \ \ \ \ \ \ \ \ \ +\ \delta_{l,l^\prime}(\delta_{m,m^\prime+1}+\delta_{m,m^\prime-1}).
\end{eqnarray}

\subsection{BCC}

The BCC state is a superposition of six plane waves ($P=6$) with
\begin{eqnarray}
&&\hat{\bf n}_1=(1,0,0),\ \ \ \ \hat{\bf n}_2=(-1,0,0),\nonumber\\
&&\hat{\bf n}_3=(0,1,0),\ \ \ \ \hat{\bf n}_4=(0,-1,0),\nonumber\\
&&\hat{\bf n}_5=(0,0,1),\ \ \ \ \hat{\bf n}_6=(0,0,-1),
\end{eqnarray}
The function $f({\bf r})$ can be expressed as
\begin{eqnarray}
f({\bf r})=2\left[\cos(2qx)+\cos(2qy)+\cos(2qz)\right],
\end{eqnarray}
which forms a 3D crystal structure. The unit cell is generated by three linearly independent vectors ${\bf a}_1=a{\bf e}_x$, ${\bf a}_2=a{\bf e}_y$, and ${\bf a}_3=a{\bf e}_z$ with the lattice spacing $a=\pi/q$. The Fourier decomposition is given by
\begin{eqnarray}
f({\bf r})=\sum_{l,m,n=-\infty}^\infty f_{lmn}e^{2iq(lx+my+nz)},
\end{eqnarray}
where the Fourier component $f_{lmn}$ reads
\begin{eqnarray}
f_{lmn}&=&\left(\delta_{l,1}+\delta_{l,-1}\right)\delta_{m,0}\delta_{n,0}
+\delta_{l,0}\left(\delta_{m,1}+\delta_{m,-1}\right)\delta_{n,0}\nonumber\\
&&\ +\ \delta_{l,0}\delta_{m,0}\left(\delta_{n,1}+\delta_{n,-1}\right).
\end{eqnarray}
The matrix form of the BdG equation is given by
\begin{eqnarray}
\sum_{l^\prime m^\prime n^\prime}{\cal H}_{[lmn],[l^\prime m^\prime n^\prime]}({\bf k})\phi_{[l^\prime m^\prime n^\prime]}({\bf k})
=E_\lambda({\bf k}) \phi_{[lmn]}({\bf k}),
\end{eqnarray}
where the Hamiltonian matrix ${\cal H}_{[lmn],[l^\prime m^\prime n^\prime]}({\bf k})$ reads
\begin{eqnarray}
\left(\begin{array}{cc} (\xi_{lmn}-\delta\mu)\delta_{l,l^\prime}\delta_{m,m^\prime}\delta_{n,n^\prime} &
\Delta f_{l-l^\prime,m-m^\prime,n-n^\prime} \\ \Delta f_{l-l^\prime,m-m^\prime,n-n^\prime}
& (-\xi_{lm}-\delta\mu)\delta_{l,l^\prime}\delta_{m,m^\prime}\delta_{n,n^\prime}  \end{array}\right)
\end{eqnarray}
with
\begin{eqnarray}
\xi_{lmn}=(k_x+2lq)^2+(k_y+2mq)^2+(k_z+2nq)^2-\mu.
\end{eqnarray}
The BZ can be defined as $k_x,k_y,k_z\in[-q,q]$. The GL coefficients $\alpha_{2k}$ ($k\geq2$) are given by
\begin{eqnarray}
\alpha_{2k}&=&\int_{-\infty}^\infty\frac{dE}{2\pi}\int_{-q}^q\frac{dk_x}{2\pi}\int_{-q}^q\frac{dk_y}{2\pi}\int_{-q}^q\frac{dk_z}{2\pi}\nonumber\\
&&{\rm Tr}_{[lmn]}\left[(S_+FS_-F)^k\right].
\end{eqnarray}
The matrix elements of $S_\pm$ and $F$ read
\begin{eqnarray}
&&(S_\pm)_{[lmn],[l^\prime m^\prime n^\prime]}=\frac{\delta_{l,l^\prime}\delta_{m,m^\prime}\delta_{n,n^\prime}}{iE+\delta\mu\mp\xi_{lmn}},\nonumber\\
&&F_{[lmn],[l^\prime m^\prime n^\prime]}=(\delta_{l,l^\prime+1}+\delta_{l,l^\prime-1})\delta_{m,m^\prime}\delta_{n,n^\prime}\nonumber\\
&&\ \ \ \ \ \ \ \ \ \ \ \ \ \ \ \ \ \ \ +\ \delta_{l,l^\prime}(\delta_{m,m^\prime+1}+\delta_{m,m^\prime-1})\delta_{n,n^\prime}\nonumber\\
&&\ \ \ \ \ \ \ \ \ \ \ \ \ \ \ \ \ \ \ +\ \delta_{l,l^\prime}\delta_{m,m^\prime}(\delta_{n,n^\prime+1}+\delta_{n,n^\prime-1}).
\end{eqnarray}

\subsection{FCC}

The FCC state is a superposition of eight plane waves ($P=8$) with
\begin{eqnarray}
&&\hat{\bf n}_1=\frac{1}{\sqrt{3}}(1,1,1),\ \ \ \ \hat{\bf n}_2=\frac{1}{\sqrt{3}}(-1,-1,-1),\nonumber\\
&&\hat{\bf n}_3=\frac{1}{\sqrt{3}}(1,-1,1),\ \ \ \ \hat{\bf n}_4=\frac{1}{\sqrt{3}}(-1,1,-1),\nonumber\\
&&\hat{\bf n}_5=\frac{1}{\sqrt{3}}(1,-1,-1),\ \ \ \ \hat{\bf n}_6=\frac{1}{\sqrt{3}}(-1,1,1),\nonumber\\
&&\hat{\bf n}_7=\frac{1}{\sqrt{3}}(1,1,-1),\ \ \ \ \hat{\bf n}_8=\frac{1}{\sqrt{3}}(-1,-1,1).
\end{eqnarray}
The function $f({\bf r})$ can be expressed as
\begin{eqnarray}
f({\bf r})=8\cos\left(\frac{2qx}{\sqrt{3}}\right)\cos\left(\frac{2qy}{\sqrt{3}}\right)\cos\left(\frac{2qz}{\sqrt{3}}\right),
\end{eqnarray}
which forms a 3D crystal structure. The unit cell is generated by three linearly independent vectors ${\bf a}_1=a{\bf e}_x$, ${\bf a}_2=a{\bf e}_y$, and ${\bf a}_3=a{\bf e}_z$ with the lattice spacing $a=\sqrt{3}\pi/q$. The Fourier decomposition is given by
\begin{eqnarray}
f({\bf r})=\sum_{l,m,n=-\infty}^\infty f_{lmn}e^{\frac{2i}{\sqrt{3}}q(lx+my+nz)},
\end{eqnarray}
where the Fourier component $f_{lmn}$ reads
\begin{eqnarray}
f_{lmn}=\left(\delta_{l,1}+\delta_{l,-1}\right)\left(\delta_{m,1}+\delta_{m,-1}\right)\left(\delta_{n,1}+\delta_{n,-1}\right).
\end{eqnarray}
The matrix form of the BdG equation and the Hamiltonian matrix ${\cal H}_{[lmn],[l^\prime m^\prime n^\prime]}({\bf k})$ take the same forms
as (104) and (105) but with $\xi_{lmn}$ given by
\begin{eqnarray}
\xi_{lmn}=\left(k_x+\frac{2lq}{\sqrt{3}}\right)^2+\left(k_y+\frac{2mq}{\sqrt{3}}\right)^2+\left(k_z+\frac{2nq}{\sqrt{3}}\right)^2-\mu.
\end{eqnarray}
The BZ can be defined as $k_x,k_y,k_z\in[-\frac{q}{\sqrt{3}},\frac{q}{\sqrt{3}}]$. The GL coefficients $\alpha_{2k}$ ($k\geq2$) are given by
\begin{eqnarray}
\alpha_{2k}&=&\int_{-\infty}^\infty\frac{dE}{2\pi}\int_{-\frac{q}{\sqrt{3}}}^\frac{q}{\sqrt{3}}\frac{dk_x}{2\pi}
\int_{-\frac{q}{\sqrt{3}}}^\frac{q}{\sqrt{3}}\frac{dk_y}{2\pi}\int_{-\frac{q}{\sqrt{3}}}^\frac{q}{\sqrt{3}}\frac{dk_z}{2\pi}\nonumber\\
&&{\rm Tr}_{[lmn]}\left[(S_+FS_-F)^k\right].
\end{eqnarray}
The matrix elements of $S_\pm$ and $F$ can be expressed as
\begin{eqnarray}
&&(S_\pm)_{[lmn],[l^\prime m^\prime n^\prime]}=\frac{\delta_{l,l^\prime}\delta_{m,m^\prime}\delta_{n,n^\prime}}{iE+\delta\mu\mp\xi_{lmn}},\nonumber\\
&&F_{[lmn],[l^\prime m^\prime n^\prime]}=(\delta_{l,l^\prime+1}+\delta_{l,l^\prime-1})(\delta_{m,m^\prime+1}+\delta_{m,m^\prime-1})\nonumber\\
&&\ \ \ \ \ \ \ \ \ \ \ \ \ \ \ \ \ \ \ \ \ \ \ \ \times\ (\delta_{n,n^\prime+1}+\delta_{n,n^\prime-1}).
\end{eqnarray}

\section{Summary}\label{s5}

In this work, we presented a new derivation of the GL free energy of crystalline (color) superconductors and compared it with the conventional diagrammatic derivation~\cite{Bowers2002}. The diagrammatic derivation and our derivation can be attributed to the use of two different representations of the fermion field: The diagrammatic derivation employs the usual continuous momentum representation, while our derivation
uses the discrete Bloch representation. In either formalism, we need to evaluate the trace of the form ${\rm Tr}
\left[(S_+F_+S_-F_-)^k\right]$. In the usual momentum representation, taking this trace leads to a summation of all possible configurations that satisfy the momentum constraint (\ref{Mconstraint}). For large $k$, the number of these configurations becomes also large, which makes the calculation tedious. In our formalism, this trace is taken in the discrete Bloch space.  Therefore, the calculation of the GL coefficients can be computerized: One can generate the matrices $F_\pm$ and $S_\pm$ and perform the matrix operations by using a computer with a proper code. Once the
computerization is realized, we may be able to evaluate the GL free energy to arbitrary order in $\Delta$. With the information of the higher-order terms in the GL free energy, we can give more reliable predictions for the phase transitions.

\emph{Acknowledgments} --- The work of G. C. was supported by the NSFC under Grant No. 11335005 and the MOST under Grant Nos. 2013CB922000 and 2014CB845400. The work of L. H. was supported by the US Department of Energy Topical Collaboration ``Neutrinos and Nucleosynthesis in Hot and Dense Matter". L. H. also acknowledges the support from Frankfurt Institute for Advanced Studies in the early stage of this work.

\end{document}